\documentclass[aip,jcp,amsmath,amssymb,reprint,nofootinbib,floatfix]{revtex4-2}

\usepackage{graphicx}
\usepackage{bm}
\usepackage{booktabs}
\usepackage[colorlinks=true,allcolors=blue]{hyperref}
\usepackage{xurl}  % long repository URL breaks across lines

\newcommand{\Nov}{N_{ov}}
\newcommand{\Naux}{N_{\mathrm{aux}}}
\newcommand{\T}{\mathsf{T}}

\begin{document}

\title{Rank-adaptive basis-update and Galerkin solvers for the random phase approximation and coupled-cluster theory}

\author{Ke Liao}
\email{ke.liao.whu@gmail.com}
\affiliation{Department of Chemistry, Yale University, New Haven, Connecticut 06520,
USA}
\author{Christian Venturella}
\affiliation{Department of Chemistry, Yale University, New Haven, Connecticut 06520,
USA}
\author{Tianyu Zhu}
\email{tianyu.zhu@yale.edu}
\affiliation{Department of Chemistry, Yale University, New Haven, Connecticut 06520,
USA}

\date{\today}

\begin{abstract}
Low-rank factorizations of coupled-cluster doubles amplitudes usually rely on
a basis fixed from perturbation theory, which may omit directions important
for the solution. We adapt the basis-update and Galerkin (BUG) scheme from
dynamical low-rank approximation to the stationary amplitude equations of the
direct random phase approximation (dRPA) and coupled cluster with single and
double excitations (CCSD). Each iteration expands the basis using the
residual, updates the amplitudes in the enlarged space, and truncates weak
modes with a relative threshold. For BUG-dRPA, a Casida construction selects
the physical solution of each projected equation, and the full update has
formal \(O(N^4)\) cost. We show that the adaptive basis closely tracks the
dominant subspace of the exact amplitudes, and the retained rank grows
approximately linearly with system size, giving cubic storage of the
amplitudes. For water clusters, BUG-CCSD gives smaller errors than fixed
perturbative bases of similar rank. Error cancellation yields accurate
interaction energies at loose thresholds for the dRPA correlation contribution
to CO adsorption on Cu(111) and for benzene dimers at the CCSD and CCSD(T)
levels.
\end{abstract}

\maketitle

\section{Introduction}

Coupled-cluster theory provides a systematic hierarchy for describing
electron correlation in molecules and solids \cite{bartlett,ccsolids}.
Coupled cluster with single and double excitations (CCSD) and its
perturbative triples extension, CCSD(T), serve as reference methods for
molecular and materials energetics \cite{purvis,raghavachari}, but their
computational cost limits the systems that can be treated.  The doubles
amplitudes are a four-index tensor whose storage grows as the fourth power
of system size, and each CCSD iteration contracts them with the
two-electron integrals at sixth-power cost.  A compact factorization of
the amplitudes reduces the storage directly.  It also lets these
contractions act on the small factors instead of the full tensor, which
lowers their cost.  Lower memory and cheaper contractions together extend
the system sizes that CCSD and its variants can reach.

To this end, several existing strategies exploit low-rank structure in the integrals or correlated
wavefunction. Density fitting (DF) and tensor hypercontraction (THC) replace
two-electron integrals with compact factors
\cite{whitten,thc1,isdf,liao_isdf_tc}, while local correlation methods
exploit the spatial structure of electron correlation \cite{localcc,dlpno}.
Global low-rank methods view the doubles amplitudes as a matrix whose rows
and columns are labeled by pairs of one occupied and one virtual orbital.
They expand this matrix in a small set of vectors in the pair space, which
we call the pair basis.  This approach began with singular-value
decompositions in approximate coupled-cluster calculations
\cite{kinoshita2003} and led to rank-reduced CCSD (RR-CCSD) methods
\cite{rrcc1,lesiuk,rrcc3}.  THC factorization of both the amplitudes and
the Coulomb integrals gives quartic-scaling CCSD \cite{rrcc3,pimpel_thc}, as does
tensor-structured coupled cluster, which solves directly for canonical
polyadic amplitude factors \cite{schutski2017}.

The effectiveness of a low-rank representation depends on the pair basis.
A common choice is to construct it from second- or third-order
M\o{}ller--Plesset amplitudes (MP2 or MP3) and hold it fixed during the CCSD
iterations \cite{rrcc1,hohenstein_gpu,lesiuk,rrcc3,pimpel_thc}.  The CCSD
iterations can adjust the amplitudes within this basis, but they cannot
recover pair directions discarded in the initial perturbative truncation.

Constructing the pair basis from MP2 or MP3 amplitudes raises a further
concern in metals, where both methods diverge in the thermodynamic limit
\cite{gellmann,shepherdmetal}.
The random phase approximation (RPA) removes this divergence
\cite{gellmann}, and natural orbitals constructed from it offer an
alternative to MP2 natural orbitals \cite{rpalno}.  The direct ring
coupled-cluster doubles (drCCD) formulation of direct RPA (dRPA) provides
the amplitudes from which these orbitals are constructed \cite{shs}.  A
dense drCCD iteration has sixth-power cost, which density fitting reduces
to fifth power and further factorization to fourth power \cite{shs,song}.
Frequency-integration algorithms reach the same fourth-power cost with
density fitting, but they give the energy without the amplitudes
\cite{eshuis}.  The drCCD equations also have unphysical solutions, to
which standard amplitude iteration can converge in small-gap systems
\cite{rekkedal,song}.  Improved preconditioners stabilize the iteration
\cite{song}, but convergence alone does not identify the physical branch.

\begin{figure*}[t!]
\includegraphics[width=\textwidth]{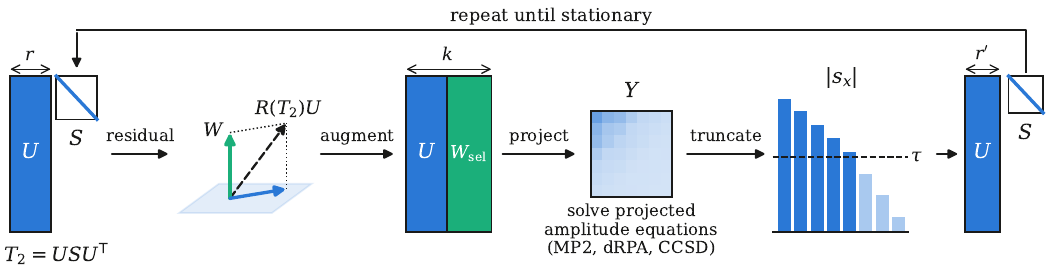}
\caption{One BUG iteration.  The residual action $R(T_2)U$ splits into
parts inside $\mathrm{span}(U)$ (blue) and outside it ($W$, green;
Eq.~\eqref{eq:residualdirections}).  In the augment step, the dominant
directions $W_{\mathrm{sel}}$ of $W$ are orthonormalized and appended to $U$,
the $k \times k$ core $Y$ is obtained from the projected amplitude equations,
and modes below the relative threshold $\tau$ are discarded, giving the
updated rank $r'$.}
\label{fig:bug}
\end{figure*}

Here we adapt the basis-update and Galerkin (BUG) approach, originally
developed for dynamical low-rank approximation \cite{dlra,projectorsplit,bug,bugadaptive}, to the
stationary MP2, dRPA and CCSD amplitude equations.  Each iteration expands
the pair basis using the residual, updates the amplitudes in the enlarged
space, and discards weak directions according to a spectral threshold.
The basis and its rank thus evolve with the amplitudes.  BUG
integrators have also been developed for matrix product states
\cite{bugmps1,bugmps2}, and the augment--project--truncate loop resembles
the residual-based enrichment of alternating minimal energy (AMEn) methods
for tensor trains \cite{amen}.  For CCSD, BUG-MP2 supplies the starting
basis, and the CCSD residual then iteratively adds pair directions that MP2 misses.
Using the resulting BUG-CCSD amplitudes, we compare the conventional (T)
correction with a version that projects all three pair indices into the
retained doubles basis.
The drCCD equation is an algebraic Riccati
equation, for which Galerkin projection onto a small subspace is
established \cite{heyouni2009,simoncini2014}.  This projection offers a
way around the unphysical solutions.  Each projected equation is small
enough to solve directly as a Casida problem, and we select its physical
branch from the positive Casida frequencies.  BUG-dRPA also yields
compressed amplitudes at $\mathcal{O}(N^4)$ cost.

\section{Theory}
\label{sec:method}
\subsection{Adaptive pair basis}
\label{sec:bugupdate}

For a closed-shell system with real orbitals, the doubles amplitude tensor is symmetric in the pair
indices $(ia)$ and $(jb)$, where $i,j$ label occupied orbitals and $a,b$
label virtual orbitals.  We represent it as
\begin{equation}
T_2=USU^{\T},\qquad U\in\mathbb{R}^{\Nov\times r},\qquad U^{\T}U=I,
\label{eq:factor}
\end{equation}
where $\Nov=N_oN_v$, with $N_o$ occupied and $N_v$ virtual orbitals.  The columns of
$U$ form an orthonormal pair basis, and the diagonal core
$S=\operatorname{diag}(s)$ contains the signed amplitude eigenvalues
retained after truncation. The rank $r$ is the number of retained pair
directions
(Fig.~\ref{fig:bug}).

Let $R(T_2)=0$ denote the MP2, dRPA or CCSD amplitude equation.
Updating the core in a fixed basis $U$ reduces the projected residual but
cannot introduce directions outside $\mathrm{span}(U)$. We obtain
expansion directions from the part of the residual action that lies
outside $\mathrm{span}(U)$ (Fig.~\ref{fig:bug}, augment step),
\begin{equation}
W=(I-UU^{\T})R(T_2)U.
\label{eq:residualdirections}
\end{equation}
Forming $R(T_2)U$ supplies candidate directions without storing the full
residual matrix. For CCSD, the implementation uses a Jacobi-preconditioned
residual action, as specified in Sec.~\ref{sec:compdetails}.

The basis-update and Galerkin (BUG) iteration \cite{bug,bugadaptive}
combines this expansion with a projected update and spectral truncation.
\begin{enumerate}\itemsep0pt
\item \textbf{Augment.} Select a block $W_{\mathrm{sel}}$ of $m$ dominant
      expansion directions from $W$, then augment the pair basis,
      $U_a=\operatorname{orth}([U,W_{\mathrm{sel}}])$
      of width $k=r+m$.
\item \textbf{Project.} Update the symmetric $k\times k$ core $Y$ in the enlarged basis
      (Fig.~\ref{fig:bug}).  The initial core
      $Y_0=(U_a^{\T}U)S(U^{\T}U_a)$ embeds the current amplitude in this
      basis.  The update targets the projected equation
      $U_a^{\T}R(U_aYU_a^{\T})U_a=0$
      (Sec.~\ref{sec:projectedequations}).
\item \textbf{Truncate.} Diagonalize $Y=Q\operatorname{diag}(s)Q^{\T}$
      and retain the modes with $|s_x|>\tau\max_y|s_y|$, where $\tau$ is
      a relative truncation threshold.  Update $U\leftarrow U_aQ_{\mathrm{keep}}$ and
      $S\leftarrow\operatorname{diag}(s_{\mathrm{keep}})$.
\end{enumerate}

Repeating these steps lets the basis change as the amplitudes approach
self-consistency.

\subsection{Projected amplitude equations}
\label{sec:projectedequations}

MP2, dRPA and CCSD share the basis enrichment and truncation.  They differ
in the projected equation for the core update.

\subsubsection{MP2}
\label{sec:mp2}

Define positive pair gaps $d_{ia}=\varepsilon_a-\varepsilon_i$, the diagonal
operator $D_{ia,jb}=d_{ia}\delta_{ij}\delta_{ab}$, and the Coulomb matrix
$V_{ia,jb}=(ia|jb)$.  The MP2 amplitude equation is
\begin{equation}
R(T_2)=V+DT_2+T_2D=0.
\label{eq:mp2residual}
\end{equation}
Substituting $T_2=U_aYU_a^{\T}$ and projecting all terms onto $\mathrm{span}(U_a)$ gives
\begin{equation}
\begin{aligned}
&EY+YE=-G,\\
&E=U_a^{\T}DU_a,\quad G=U_a^{\T}VU_a.
\end{aligned}
\label{eq:mp2core}
\end{equation}
This linear Sylvester equation \cite{bartelsstewart,simoncini} has the same
form as the MP2 amplitude equations in noncanonical orbitals
\cite{pulaysaebo}. We keep the off-diagonal elements of the projected gap matrix $E$, which is
generally not diagonal in the BUG basis, and diagonalize it,
$E=P\operatorname{diag}(\epsilon)P^{\T}$.  In this eigenbasis the core
decouples as
$\widetilde Y_{\mu\nu}=-\widetilde G_{\mu\nu}/(\epsilon_\mu+\epsilon_\nu)$,
with $\widetilde Y=P^{\T}YP$ and $\widetilde G=P^{\T}GP$, at $O(k^3)$ cost.
The closed-shell correlation energy is
$E_c=\sum_{ijab}(T_2)_{ia,jb}[2(ia|jb)-(ib|ja)]$.
We call the energy with only the first (direct) term direct MP2 and the
full expression full MP2; both use the amplitudes of Eq.~\eqref{eq:mp2core}.
The corresponding direct and exchange contractions are given in
Sec.~S4\,A of the Supplementary Material.

\subsubsection{Direct RPA}
\label{sec:drpa}

In the drCCD formulation of closed-shell dRPA \cite{shs}, the amplitude
equation is
\begin{equation}
\begin{aligned}
2V+AT_2+T_2A+2T_2VT_2&=0,\\
A&=D+2V,
\end{aligned}
\label{eq:care}
\end{equation}
with correlation energy $E_c=\operatorname{Tr}(VT_2)$.  Projection onto the
enlarged basis gives
\begin{equation}
\begin{aligned}
&\widetilde V+EY+YE+Y\widetilde VY=0,\\
&E=U_a^{\T}AU_a,\quad\widetilde V=2U_a^{\T}VU_a.
\end{aligned}
\label{eq:projectedcare}
\end{equation}
This Riccati equation has several algebraic solutions, as do other
projected Riccati equations \cite{heyouni2009,simoncini2014}.  The physical
solution uses positive-frequency RPA excitations and reproduces the
frequency-integration energy \cite{shs}, whereas the others replace some
excitations with their negative-frequency partners.  For positive
gaps and a stable reference, the Casida construction selects the physical
solution. Since $A=D+2V$, the projected gap matrix is
$E-\widetilde V=U_a^{\T}DU_a$. We compute its Cholesky factorization
$E-\widetilde V=LL^{\T}$ and diagonalize the symmetric $k\times k$ matrix
$L^{\T}(E+\widetilde V)L$.  Its eigenvalues are the squared frequencies
$\omega^2$.  Taking $\omega>0$ and reconstructing the corresponding Casida
eigenvectors yields the physical core $Y$, as detailed in the Supplementary
Material.
This core solve costs $\mathcal O(k^3)$, compared with
$\mathcal O(\Nov^3)$ for the full pair-space Casida problem.  Because the
full update applies $V$ only through its density-fitting factors to
$U_a$, it requires $O(N^4)$ work (Sec.~\ref{sec:cost}).
The resulting amplitude eigenvalues satisfy $|s_x|<1$, giving the inexpensive
diagnostic $\lambda_{\max}(T_2^{\T}T_2)=\max_xs_x^2<1$ \cite{song}.

Metallic systems are treated with a smeared reference and fractional
orbital occupations $f_p\in[0,2]$.  Active transitions $p\to q$ then have
$f_p>f_q$ and $\varepsilon_q>\varepsilon_p$.
The density-fitting Coulomb factors $B_{pq,Q}$, with auxiliary index $Q$,
represent these transitions. Their weighted form
$\widetilde B_{pq,Q}=\sqrt{n_{pq}}B_{pq,Q}$, where
$n_{pq}=(f_p-f_q)/2$, defines the Coulomb matrix in the active transition
space.  The diagonal entries of $D$ remain the orbital-energy differences
$\varepsilon_q-\varepsilon_p$.  The active window and treatment of inactive
pairs are specified in the Supplementary Material.  Both BUG and the
frequency-integration reference evaluate this occupation-weighted
zero-temperature dRPA expression for the smeared reference.
% RESOLVED [A03]: Production cutoff n=(fp-fq)/2 > 1e-6 confirmed in
% low_scaling_rpa at 33ad0b2. The SI defines the active domain, zero-weight
% pairs, positive-gap condition and reference-window differences.

\subsubsection{CCSD}
\label{sec:ccsd}

CCSD requires the singles and doubles projections of the
similarity-transformed Hamiltonian to vanish \cite{hirata}.  Separating the
orbital-gap terms from the doubles residual defines the remaining
numerator $N(T_1,T_2)$,
\begin{equation}
R_2(T_1,T_2)=DT_2+T_2D+N(T_1,T_2).
\label{eq:ccsdnumerator}
\end{equation}
The projected doubles equation is
\begin{equation}
\begin{aligned}
EY+YE&=-U_a^{\T}N(T_1,U_aYU_a^{\T})U_a,\\
E&=U_a^{\T}DU_a.
\end{aligned}
\label{eq:ccsdcore}
\end{equation}
Unlike the MP2 numerator, $N$ depends nonlinearly on the amplitudes.
Density-fitting contractions apply $N(T_1,T_2)$ to basis blocks using the
factors $U$ and $S$.
BUG-MP2 iterations provide the initial pair subspace.

At each outer iteration, we evaluate $N(T_1,T_2)$ using the current singles
amplitudes $T_1$ and doubles amplitudes $T_2=USU^{\T}$.  We hold the
projected numerator $G_a=U_a^{\T}N(T_1,T_2)U_a$ fixed for one undamped
Sylvester update, $EY+YE=-G_a$.  $G_a$ is assembled from blocks.
With the selected directions $W_{\mathrm{sel}}$ orthonormalized against $U$
and among themselves, the enlarged basis is $U_a=[U,W_{\mathrm{sel}}]$ and
$G_a$ has the blocks
$U^{\T}NU$, $U^{\T}NW_{\mathrm{sel}}$ and
$W_{\mathrm{sel}}^{\T}NW_{\mathrm{sel}}$, with the fourth given by symmetry.
The first two follow from $NU$, already computed to form the residual
action $R_2U$ for basis enrichment. Only
$W_{\mathrm{sel}}^{\T}NW_{\mathrm{sel}}$ requires a new action of $N$,
on the $m$ added directions.
We then diagonalize and truncate the updated core $Y$ as in
Sec.~\ref{sec:bugupdate} and take one Jacobi step for the singles amplitudes.

\subsubsection{Perturbative triples on BUG amplitudes}
\label{sec:triples}

After the BUG-CCSD iterations meet the stopping criterion, we reconstruct
the doubles amplitudes from Eq.~\eqref{eq:factor} and evaluate the
conventional noniterative (T) correction \cite{raghavachari} with the BUG
singles and doubles. The triples numerators and orbital-energy denominators
are evaluated in the saved canonical orbital basis, with no further
projection or truncation. As $\tau\to0$, the BUG-CCSD amplitudes approach
the canonical CCSD amplitudes, and this correction reduces to canonical (T).

As an alternative, projected BUG--(T) instead restricts all three
occupied--virtual pair indices of the triples to the retained doubles basis,
lowering the formal cost from $O(N^7)$ to $O(N^6)$ (see Sec.~S5 of the
Supplementary Material for detailed formulation).

\subsection{Spectral truncation and energy error}
\label{sec:truncation}
Let $T_{2,a}=U_aYU_a^{\T}$ be the updated amplitude in the enlarged basis
(Sec.~\ref{sec:bugupdate}).  By the Eckart--Young theorem \cite{eckartyoung},
spectral truncation gives the best rank-$r$ approximation to $T_{2,a}$ in
the Frobenius norm.
For symmetric $Y$, the singular values are $|s_x|$, so retaining the largest
magnitudes minimizes the squared error,
$\|Y-Y_r\|_F^2=\sum_{\mathrm{disc}}s_x^2$.
Let $T_{2,a,r}$ be the rank-$r$ truncation of $T_{2,a}$, and $T_{2,r}^\ast$ the
best rank-$r$ approximation of the exact amplitude $T_2^\ast$.  The triangle
inequality together with the Eckart--Young property gives
\begin{equation}
\|T_2^\ast-T_{2,a,r}\|_F\le\|T_2^\ast-T_{2,r}^\ast\|_F+2\|T_{2,a}-T_2^\ast\|_F
\label{eq:trackingbound}
\end{equation}
(Sec.~S1\,B of the Supplementary Material).  The excess error over the best
rank-$r$ approximation is therefore bounded
by twice the error of the updated amplitude.

How much energy does truncation lose?  For dRPA this has an exact answer.
Write the exact amplitude in its eigenmodes,
$T_2^\ast=\sum_xs_xu_xu_x^{\T}$, and let $T_{2,r}^\ast$ keep only the $r$ modes
with the largest $|s_x|$.  The dRPA energy $\operatorname{Tr}(VT_2)$ is
linear in the amplitude, so each discarded mode removes
$s_x\langle u_x|V|u_x\rangle$.  The amplitude equation, taken in the same
mode, ties this Coulomb term to $s_x$ itself (Supplementary Material),
and the energy change becomes
\begin{equation}
\Delta E=\sum_{\mathrm{disc}}s_x^2g_x,\qquad
g_x=\frac{\langle u_x|D+2V|u_x\rangle}{1+s_x^2},
\label{eq:truncerror}
\end{equation}
where $\Delta E=E_c[T_{2,r}^\ast]-E_c[T_2^\ast]$.  For positive orbital gaps and
a stable reference, every weight $g_x$ is positive.  Each discarded mode
therefore raises the energy by an amount proportional to the square of its
eigenvalue, much as the discarded density-matrix weight controls the error
in the density-matrix renormalization group \cite{white}.  Direct MP2 obeys an identity of the same form
(Supplementary Material).

\begin{figure*}[t!]
\includegraphics[width=\textwidth]{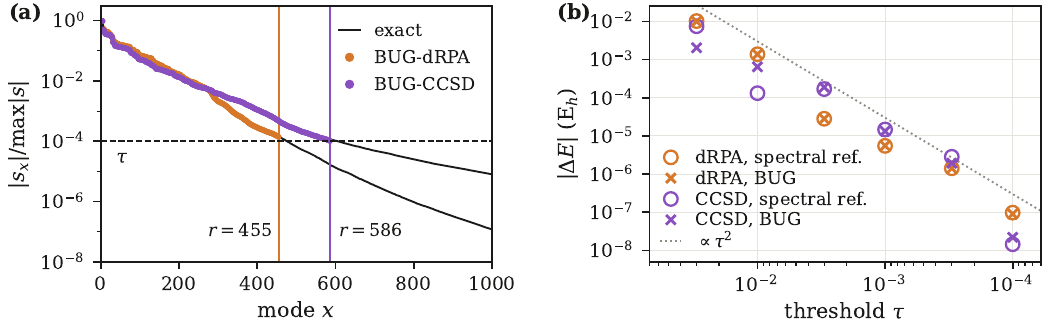}
\caption{Comparison of BUG with exact spectral truncation for (H$_2$O)$_5$
in the cc-pVDZ basis, with all electrons correlated.
(a) Exact dRPA and CCSD amplitude spectra (lines) and
BUG core eigenvalues at $\tau=10^{-4}$ (dots), for the first 1000 of 2375 modes.
Vertical lines mark the retained ranks.  (b) Correlation-energy errors of BUG--dRPA and
BUG--CCSD (crosses) and of the exact amplitudes truncated to the same rank
(circles); for CCSD the exact singles are kept.  The dotted line is
$\propto\tau^2$.}
\label{fig:basisquality}
\end{figure*}

The eigenvalues decay roughly exponentially,
$|s_x|\approx s_{\max}e^{-\alpha x}$
[Fig.~\ref{fig:basisquality}(a)].  If the weights vary slowly near the
cutoff, discarding all modes with $|s_x|\le\tau s_{\max}$ gives
\begin{equation}
\Delta E\approx\bar g\,\frac{s_{\max}^2}{2\alpha}\,\tau^2,
\label{eq:tausquared}
\end{equation}
where $\bar g$ is the average weight near the cutoff.  Lowering $\tau$ by
a factor of ten thus lowers the error by about a factor of one hundred.  For
$N$ non-interacting identical units the spectrum is that of one unit
repeated $N$ times, so at fixed $\tau$ both the rank and $\Delta E$ are
proportional to $N$, and the relative error does not depend on size.  The
CCSD energy has no such identity, and its threshold dependence is assessed
numerically in Sec.~\ref{sec:basisbenchmark}.

\subsection{Computational cost}
\label{sec:cost}
\label{sec:scaling}

The retained rank is found to grow linearly with system size
(Sec.~\ref{sec:resultsall}).  With
$N_o\sim N_v\sim\Naux\sim r\sim k\sim N$ and $\Nov=N_oN_v\sim N^2$, where
$\Naux$ is the number of auxiliary functions and $k$ the enlarged trial rank,
a BUG update costs $O(N^4)$ for direct MP2 and dRPA.  Evaluating the full-MP2 exchange
energy costs $O(N^5)$, as in canonical DF-MP2.  CCSD costs $O(N^6)$,
because its quadratic exchange terms contract two amplitude factors whose
modes are not paired, and conventional (T) costs $O(N^7)$.  The projected
cores add only $O(k^3)$ solves.  Section~S4 of the Supplementary Material
gives the term-level counts.

For a fixed system, enlarging the orbital basis keeps $N_o$ fixed while $N_v$
and $\Naux$ grow. The retained rank grows much more slowly than $N_v$, allowing
savings that the system-size scaling does not capture.  In large basis sets
the particle--particle ladder dominates canonical
DF-CCSD.  It assembles the four-virtual integrals from density-fitting
factors and contracts them with the amplitudes, at $O(\Naux N_v^4)$ and
$O(N_o^2N_v^4)$ cost per iteration.  BUG-CCSD instead applies the
density-fitting factors to the amplitude factors one virtual index at a
time, at $O(N_oN_v\Naux r(N_v+k))$ cost.  The ring terms cost $O(N_o^3N_v^3)$ in both
implementations.  The benzene-dimer timings in Fig.~\ref{fig:timing}(c)
illustrate these
savings.

The doubles are stored as $U$ and $s$, which require $O(N^3)$ storage
instead of the $O(N^4)$ of a dense amplitude tensor, and the MP2 and dRPA
solvers keep all intermediates cubic.  The present CCSD code also builds a
transient $O(N^4)$ intermediate that collects the ladder and crossed-ring
terms for efficient matrix multiplication.  This is an implementation
choice and is not required by BUG (Sec.~\ref{sec:timing}).

Because the orbital-energy denominator is resolved inside the small core,
the remaining contractions are purely spatial and can act on $U$ one factor
at a time.  Separable interaction factors from interpolative separable density
fitting (ISDF) or THC \cite{thc1,isdf,liao_isdf_tc} then give a route to full MP2 in $O(N^4)$ time
and $O(N^3)$ memory, comparable to $O(N^4\log N)$ plane-wave MP2
\cite{schaefer}, and to CCSD in $O(N^5)$ time and $O(N^3)$ memory. Section~S4 of the Supplementary Material outlines these routes, which are
left for future work.

\section{Computational Details}
\label{sec:compdetails}

The BUG solvers are implemented in \texttt{pytc} \cite{pytc,liao_isdf_tc},
using PySCF 2.11.0 \cite{pyscf,pyscf2018,pyscf2026}, NumPy 2.3.3 and JAX/jaxlib 0.8.1
in 64-bit precision.  Unless stated otherwise, calculations use density
fitting and the relative truncation rule of Sec.~\ref{sec:bugupdate}, with
energy thresholds of $10^{-10}\,E_h$ for restricted Hartree--Fock (RHF)
and $10^{-9}\,E_h$ for canonical DF-CCSD.

BUG-MP2 starts from eight random orthonormal columns (seed 0) and a zero
core. The default CCSD initialization uses at most five BUG-MP2 iterations
at the chosen threshold, with zero singles.  CCSD enrichment uses the
Jacobi-preconditioned residual action, retaining all external directions
during warm-up and up to 32 leading singular vectors thereafter.  CCSD stops
when the absolute amplitude step falls below $10^{-10}$, or when
$|\Delta E_c|<\tau^2|E_c|$ holds for three consecutive iterations.
At termination, the projected doubles and singles residuals remain of order
$10^{-5}$ to $10^{-4}\,E_h$ for the water
clusters (Sec.~S12 of the Supplementary Material).  The reported CCSD
energies therefore contain an iteration error in addition to the
truncation error.  Section~S14 of the Supplementary Material gives the
complete initialization, enrichment and stopping protocols.  Small
regression checks against PySCF and fcdmft \cite{zhuchan,zhu2020fcdmft,zhu2021fullcell}
are collected in Sec.~S9.

CPU CCSD timing comparisons use 48 threads on one node with two
24-core Xeon Gold 6442Y processors, for BUG and canonical calculations alike.
BUG timings use the NumPy/BLAS implementation archived with the supporting
data.  Canonical CCSD uses
PySCF DF-RCCSD with DIIS.  These timing runs are separate from the accuracy
series.  Their stopping protocol and implementation settings are specified
in Sec.~S8.A of the Supplementary Material.

\begin{figure*}[t]
\includegraphics[width=\textwidth]{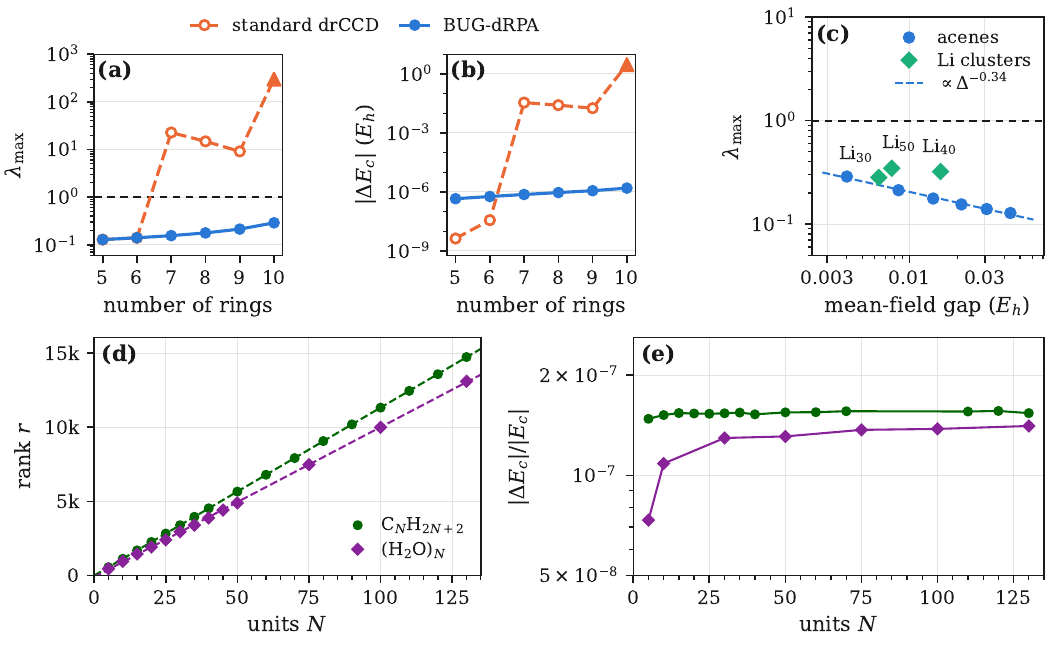}
\caption{Branch selection and accuracy of BUG-dRPA in molecules (cc-pVDZ, $\tau=10^{-4}$).  (a) Branch diagnostic
$\lambda_{\max}=\max_xs_x^2$ and (b) error against the frequency-form
dRPA reference along the acene series of Ref.~\onlinecite{song}, for the
standard drCCD iteration and BUG.  Triangles mark off-scale values where drCCD diverges.
(c) BUG-dRPA branch diagnostic versus the PBE HOMO--LUMO gap for acenes
and lithium clusters.  The power-law fit uses only the acene data.  (d) Retained
rank with linear fits and (e) relative energy error for alkanes
and water clusters.}
\label{fig:rpamolecular}
\label{fig:branch}
\label{fig:rankscaling}
\end{figure*}

\section{Results}
\label{sec:resultsall}

\subsection{Accuracy of the adaptive pair basis}
\label{sec:basisquality}
\label{sec:basisbenchmark}

To test whether the adaptive basis finds the best rank-$r$ representation,
we compare BUG for (H$_2$O)$_5$ with exact spectral truncation, i.e., the
exact amplitudes truncated to the same rank. For dRPA, the BUG energy error closely matches
that of exact spectral truncation over five orders of magnitude
[Fig.~\ref{fig:basisquality}(b)].  For $\tau\le3\times10^{-3}$ the two differ
by less than $2\times10^{-6}$ of the correlation energy.  BUG is in some
cases slightly more accurate, because spectral truncation is optimal in
norm, not in energy \cite{eckartyoung}.  The fitted log--log slope is 2.0, as
Eq.~\eqref{eq:tausquared} predicts.  For CCSD, the BUG and spectral-truncation
energy errors are comparable for $\tau\le3\times10^{-3}$.  The BUG error is
about five times larger at $\tau=10^{-2}$ and about one quarter of the
spectral-truncation error at $\tau=3\times10^{-2}$.  The CCSD errors also fall rapidly with
$\tau$, but the fitted slope is 1.9 over all six thresholds and 1.5 when the
smallest threshold, where the error changes sign, is excluded.  Direct MP2
behaves like dRPA, whereas full MP2, which includes exchange, converges more
slowly (Sec.~S1 of the Supplementary Material).

For dRPA and CCSD, the BUG core eigenvalues track the leading exact doubles
eigenvalues [Fig.~\ref{fig:basisquality}(a)].  For CCSD, the core eigenvalues agree with the leading
exact eigenvalues to a median of 0.1\% at $\tau=3\times10^{-3}$.  For CCSD as
for dRPA, the amplitude weight missed by the BUG basis is within 25\% of the
minimum possible at each rank.  For $\tau\le3\times10^{-3}$, the median
principal angle between the BUG basis and the leading exact eigenvectors is
below $0.05^\circ$ for dRPA and below $0.2^\circ$ for CCSD.  A few
weak directions near the cutoff remain poorly aligned
(Sec.~S2 of the Supplementary Material).

\subsection{dRPA in molecular systems}
\label{sec:smallgap}
\label{sec:branch}
\label{sec:results}

We first examine whether the projected dRPA solve selects the physical
branch as the gap decreases along an acene series.  We then use alkane
chains and water clusters to assess retained rank and energy accuracy
over larger size ranges.

Molecular dRPA calculations use the cc-pVDZ basis and the cc-pVDZ-JKFIT
auxiliary basis.  Alkanes
and water clusters up to (H$_2$O)$_{130}$ use RHF orbitals and
all-electron correlation; water clusters up to 50 molecules are those of
Ref.~\onlinecite{hohenstein_gpu}, and those with 75, 100 and 130 molecules
are spheres cut from a box of liquid water equilibrated at 300~K with the
simple point charge (SPC) model~\cite{berendsen_spc}, used without further
relaxation.
Acenes and lithium clusters follow
Ref.~\onlinecite{song}, using PBE orbitals~\cite{perdew1996pbe} and frozen heavy-atom 1s
orbitals.  For alkanes and water clusters, we obtain reference dRPA energies
by imaginary-frequency integration in fcdmft using modified Gauss--Legendre
quadrature~\cite{zhuchan}.  For acenes and lithium clusters, we obtain them
by diagonalizing the full dRPA response matrix.  The 130-water
inputs were generated with GPU4PySCF
\cite{gpu4pyscf_li,gpu4pyscf_wu}.  Sections~S13 and S14 of the Supplementary Material give the geometries,
hardware, iteration settings and reference checks.

Figure~\ref{fig:branch}(a,b) compares the standard MP2-preconditioned drCCD
iteration with the projected solve for the acene series of
Ref.~\onlinecite{song}.  The diagnostic $\lambda_{\max}<1$ proposed there
and defined in Sec.~\ref{sec:drpa} identifies the physical branch.  For
the smallest acenes the two solvers agree.  As the gap
closes, the standard iteration first converges to unphysical solutions with
$\lambda_{\max}\gg1$ and correlation energies below reference values,
then stalls, and finally diverges.  The energy offsets are
consistent with flipping the sign of the lowest RPA mode
\cite{song}.  The projected solve reproduces the reference dRPA energies
across the whole series with $\lambda_{\max}<1$.  For the acenes, $\lambda_{\max}$
increases as the PBE HOMO--LUMO gap narrows [Fig.~\ref{fig:branch}(c)].
The three lithium clusters lie above the acene fit and show no clear trend
with gap.

At $\tau=10^{-4}$, the retained rank grows linearly with system size for
alkane chains and water clusters [Fig.~\ref{fig:rankscaling}(d,e)] while the pair dimension
grows quadratically, so the retained fraction falls to about 1\% for the
largest systems.  Water clusters retain slightly fewer vectors
per molecule than alkanes per carbon at the same basis size per unit.
The relative correlation-energy error stays at the
$10^{-7}$ level throughout.  A fixed threshold therefore maintains nearly
constant relative accuracy
over the tested size range.

\subsection{dRPA for CO adsorption on Cu(111)}
\label{sec:cocu}

\begin{figure*}[t!]
\includegraphics[width=\textwidth]{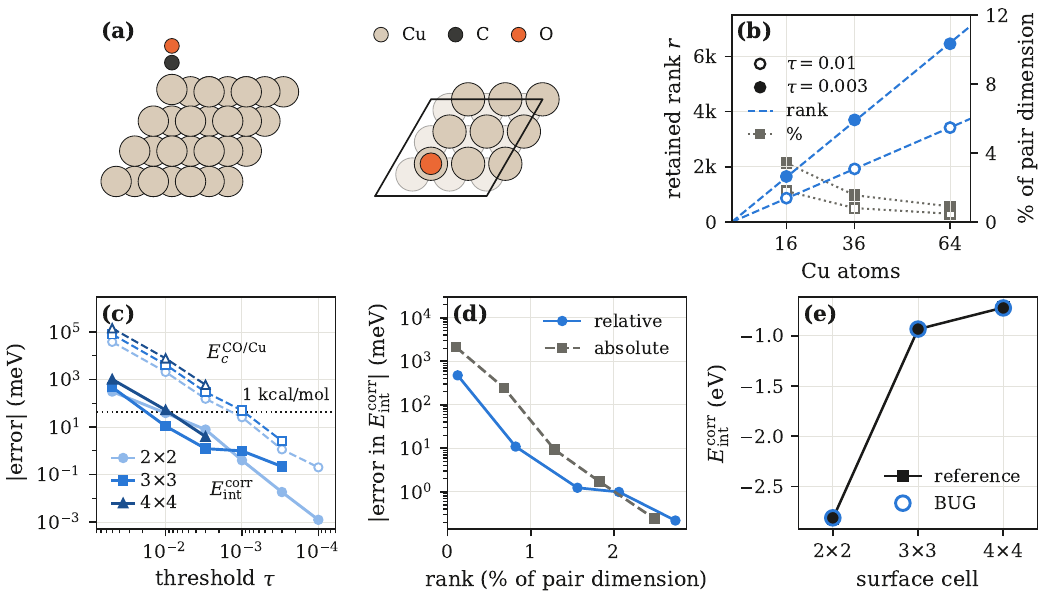}
\caption{BUG-dRPA for CO atop Cu(111) at the $\Gamma$ point (MCAO-cc-pVDZ,
GTH-PBE).  (a) The $3\times3$ cell.  (b) Clean-slab retained rank against the
number of Cu atoms at two thresholds, with proportional fits; the right axis
gives the fraction of the pair dimension.  (c) Errors in the correlation
contribution to the interaction energy $E_{\rm int}^{\rm corr}$ (solid)
and in the CO/Cu correlation energy $E_c^{\rm CO/Cu}$
(dashed) for the three cells; the dotted line marks 1~kcal/mol.  (d) Error of
$E_{\rm int}^{\rm corr}$ against retained rank for relative and absolute truncation
($3\times3$ cell).  (e) $E_{\rm int}^{\rm corr}$ from the frequency-integration reference and from BUG at
$\tau=3\times10^{-3}$.}
\label{fig:cocu}
\end{figure*}

We use CO adsorption on Cu(111) to test amplitude compression in a
small-gap system.  The correlation contribution to the interaction energy
is a small difference between much larger correlation energies.
We consider one CO molecule adsorbed atop a four-layer slab in
periodic $2\times2$, $3\times3$, and $4\times4$ surface
cells containing 16, 36, and 64 Cu atoms, respectively
[Fig.~\ref{fig:cocu}(a)].  The cells are constructed from the relaxed
$2\times2$ structure of Ref.~\onlinecite{mcao}.  The corresponding coverages are
$\tfrac14$, $\tfrac19$, and $\tfrac1{16}$ of a monolayer.  The pair
dimension grows from about 50,000 to about 700,000 across these cells.
We use the MCAO-cc-pVDZ basis of Yu et~al.\cite{mcao} and a
GTH-PBE pseudopotential that retains 19 valence electrons per Cu atom.
The PBE reference uses Fermi smearing with $\sigma=0.005\,E_h$.

We define the correlation contribution to the interaction energy as
\begin{equation}
E_{\rm int}^{\rm corr}=E_c^{\rm CO/Cu}-E_c^{\rm Cu}-E_c^{\rm CO},
\label{eq:interactionenergy}
\end{equation}
with all three terms evaluated in the same cell and atom-centered basis,
using ghost atoms for the fragments.  Errors in $E_{\rm int}^{\rm corr}$ are measured
against frequency-integration dRPA for the same reference states.
We compare the usual relative rule, $|s_i|>\tau\max_j|s_j|$, with an
absolute rule, $|s_i|>\tau$ [Fig.~\ref{fig:cocu}(d)].  The relative rule
gives smaller errors in $E_{\rm int}^{\rm corr}$ at the same CO/Cu rank.  Because the
largest amplitude is smaller for isolated CO than for the slabs, relative
normalization retains weaker molecular modes and balances the accuracy of
the three energies that are subtracted.  A common absolute cutoff is
additive over separated fragments, whereas the relative rule is not when
their largest amplitudes differ.  Except for this comparison, we use relative truncation
throughout.
% DOCUMENTED [A12]: SI tables give component ranks, core scales, cutoffs,
% stopping reasons and iteration counts from the versioned data. These are
% separate adaptive runs, not a fixed-core normalization advantage. Additional
% residual diagnostics or matched-initialization tests are not claimed.

\begin{figure*}[t!]
\includegraphics[width=\textwidth]{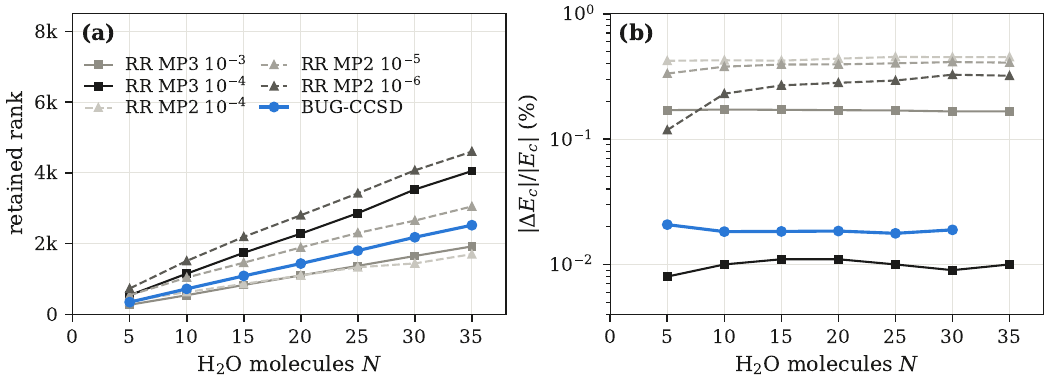}
\caption{BUG-CCSD for water clusters (TZVP, frozen O~1s, $\tau=3\times10^{-3}$).
(a) Retained rank and (b) relative correlation-energy error, compared with
RR-CCSD results digitized from Ref.~\onlinecite{hohenstein_gpu}; legend values
are the truncation thresholds of Ref.~\onlinecite{hohenstein_gpu}.
No DF-CCSD reference is available for benchmarking BUG energy error at $N=35$.}
\label{fig:ccsdwater}
\end{figure*}

At fixed threshold, the clean-slab rank is proportional to the number of Cu
atoms [Fig.~\ref{fig:cocu}(b)], so the retained fraction decreases as the
cell grows, and all calculations satisfy the physical-branch diagnostic.
The component errors largely cancel.  The error in $E_{\rm int}^{\rm corr}$ is one
to two orders of magnitude smaller than the CO/Cu correlation-energy error
and is similar across the three cells [Fig.~\ref{fig:cocu}(c)].  At
$\tau=3\times10^{-3}$, BUG reproduces the reference $E_{\rm int}^{\rm corr}$
to within 10~meV in every cell.  The reference value changes by about
1.9~eV from the $2\times2$ to the
$3\times3$ cell [Fig.~\ref{fig:cocu}(e)].  That change reflects the
finite slab-size error due to the $\Gamma$-point sampling of these small
metallic cells and the difference in CO coverage.

\subsection{BUG-CCSD with adaptive versus fixed pair bases}
\label{sec:ccsdwater}

\begin{figure*}[t]
\includegraphics[width=\textwidth]{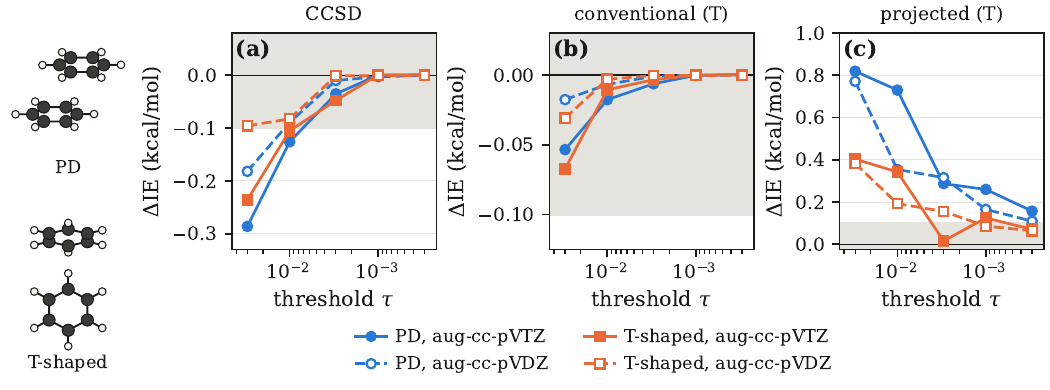}
\caption{Benzene dimers (S22 parallel-displaced, PD, and T-shaped) \cite{s22}.
Errors are BUG minus canonical DF values in the same basis
(aug-cc-pVTZ solid, aug-cc-pVDZ dashed).
Panel (a) shows the error in the counterpoise-corrected CCSD interaction energy.
Panels (b) and (c) show the error in the (T) contribution from the conventional
and projected corrections, respectively.
The shaded band marks $\pm0.1$~kcal/mol.}
\label{fig:benzene}
\end{figure*}

We next turn to BUG-CCSD for water clusters and compare its retained ranks
and correlation-energy errors with those reported in Ref.~\onlinecite{hohenstein_gpu}
for RR-CCSD with fixed perturbative pair bases. The BUG accuracy series
uses the same \texttt{jgauss-tzp1} TZVP file, runs on one H200 GPU, and
stops on the energy criterion.

The retained BUG rank grows linearly with the number of molecules, so its
share of the pair dimension falls from 13\% at 5 waters to 2\% at 35
[Fig.~\ref{fig:ccsdwater}(a)]. For clusters of 5--30 waters, the BUG
correlation-energy error relative to canonical DF-CCSD stays near 0.02\%,
below the published errors for fixed MP2 and MP3 pair bases of similar rank
[Fig.~\ref{fig:ccsdwater}(b)]. The tightest published MP3 basis reaches
about half the BUG error at 1.6 times the rank. The two studies use different
implementations and auxiliary basis sets, so the difference in accuracy at
similar ranks cannot be attributed solely to adapting the pair basis.

\subsection{Benzene-dimer interaction energies and triples}
\label{sec:triplesresults}

\begin{figure*}[t]
\includegraphics[width=\textwidth]{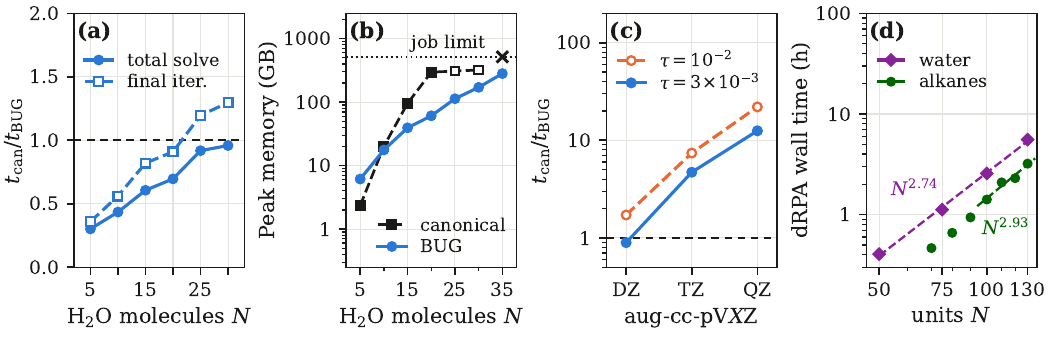}
\caption{Timing and memory benchmarks for BUG-CCSD and BUG-dRPA.
(a) Wall time of canonical DF-CCSD over that of
BUG-CCSD for water clusters (TZVP, $\tau=3\times10^{-3}$) under a shared
stopping rule, for the total solve and for one iteration at the final rank.
(b) Peak memory of the same runs.  Open squares are canonical runs with a
lower memory limit, and the cross marks the canonical run that exceeded the
515~GB job limit (dotted line).  (c) Total CCSD wall-time ratio for the
parallel-displaced benzene dimer, summing the dimer and both ghost monomers.
(d) BUG-dRPA solver wall time
on one H200 GPU with power-law fits.  All times exclude the
self-consistent-field (SCF) calculation.}
\label{fig:timing}
\end{figure*}

We then use two benzene-dimer geometries to test how compressing the doubles
amplitudes affects both the CCSD interaction energy and its perturbative
triples correction. For the parallel-displaced and T-shaped geometries from
S22 \cite{s22}, we use aug-cc-pVDZ and aug-cc-pVTZ with the corresponding
JKFIT auxiliary bases and freeze the C~1s orbitals.  We use the PySCF
DF-RCCSD(T) kernel to evaluate the conventional (T) correction with reconstructed
BUG amplitudes and compare it with the correction obtained from canonical
DF-CCSD amplitudes. Both calculations use the same orbital and auxiliary
bases and frozen-core convention.

Interaction energies are counterpoise corrected \cite{boysbernardi}.
The conventional (T) error from using BUG amplitudes is a small fraction
of the CCSD error [Fig.~\ref{fig:benzene}(a,b)]. At
$\tau=3\times10^{-3}$, the total CCSD(T) interaction energy is within
about 0.05~kcal/mol of the canonical value for both geometries in
aug-cc-pVDZ and aug-cc-pVTZ (Table~S2 of the Supplementary Material).  As for
CO/Cu, this
accuracy relies on cancellation.  The errors of the dimer and monomer
energies are several times larger than that of their difference, although
only a few percent of the dimer pair dimension is retained.

Projecting all three pair indices of the (T) correction into the retained
doubles basis reduces its formal cost from $O(N^7)$ to $O(N^6)$.
With the same BUG-CCSD amplitudes, the
projected (T) correction gives a larger interaction-energy error than the
conventional (T) correction at each common threshold [Fig.~\ref{fig:benzene}(b,c)].
At $\tau=3\times10^{-3}$, the projected (T) contribution underbinds the
parallel-displaced dimer by about 0.3~kcal/mol in both basis sets and the
T-shaped dimer by 0.16~kcal/mol in aug-cc-pVDZ and 0.02~kcal/mol in aug-cc-pVTZ.
Here, a basis adapted to the
doubles represents the triples less
accurately, supporting the use of a separate triples subspace
\cite{lesiuk_t,rickert_svd_t}.
We refer the readers to the detailed projected (T) formalism and associated
analysis on the water dimer in Sec.~S5 of the Supplementary Material.

\subsection{Solver timings}
\label{sec:timing}

Finally, we show the BUG solver timings in Figure~\ref{fig:timing}.
For water clusters, we compare BUG-CCSD at $\tau=3\times10^{-3}$ with canonical
DF-CCSD using the same energy stopping rule [Fig.~\ref{fig:timing}(a)].  BUG is slower for
small clusters and reaches parity near 30 waters.  Every completed
canonical run with DIIS passes the stopping rule at cycle 10, while undamped
BUG takes 13 to 17 iterations, the first of which are cheap because the rank
is still growing.  At the final rank a BUG iteration is cheaper than a
canonical cycle from 25 waters on, by a factor of 1.3 at 30 waters.  Both
implementations retain formal $O(N^6)$ work.

Peak memory is lower for BUG from 10 waters on, with 171~GB against
321~GB for the canonical run at 30 waters [Fig.~\ref{fig:timing}(b)].  The
canonical run exceeded the available memory at 35 waters, where BUG needs
281~GB.  The transient $O(N^4)$ intermediate of the present CCSD code
(Sec.~\ref{sec:cost}) accounts for 23\% of the BUG peak at 20 waters and
46\% at 35.  The cubic terms thus dominate over the measured range, and the
quartic one would take over in larger clusters.

For the benzene dimer, the savings grow with the orbital basis
[Fig.~\ref{fig:timing}(c)].  At $\tau=3\times10^{-3}$, BUG-CCSD is slightly
slower than canonical CCSD in aug-cc-pVDZ but about $5\times$ faster in
aug-cc-pVTZ and about $13\times$ faster in aug-cc-pVQZ, with CCSD
interaction-energy errors below 0.04~kcal/mol.  The
canonical runs, stopped by the same rule, are more accurate. Over the same
basis range, $N_v$ grows 4.3-fold, while the retained dimer rank grows only
from 817 to 1467.

For the parallel-displaced benzene dimer, we time the projected and
canonical (T) corrections using precomputed CCSD amplitudes.
Both methods use 48 threads and 515~GB of memory on nodes with two
24-core Xeon Gold 6442Y processors.  In
aug-cc-pVTZ at $\tau=10^{-3}$, the projected correction takes 6.3~h and
the canonical one 8.6~h.  In aug-cc-pVQZ, canonical (T) takes 100~h, while
the projected correction takes 6.2~h at $\tau=3\times10^{-3}$ and 24.8~h
at $\tau=10^{-3}$.  These projected corrections recover 96\%, 91\% and
94\% of the canonical (T) contribution to the dimer correlation energy,
respectively.  Table~S5 of the Supplementary
Material lists the timings.

On one H200 GPU, the fitted BUG-dRPA timing exponents are 2.9 for
C$_{100}$--C$_{130}$ and 2.7 for clusters of 50--130 waters
[Fig.~\ref{fig:timing}(d)].  For 130 waters, BUG-dRPA yields the correlation
energy and compressed drCCD amplitudes on the physical branch in 5.6~h on
one H200 GPU.  The energy-only frequency-integration reference takes 4.3~h
on one 48-core CPU node (Sec.~S14\,C of the Supplementary Material).  Section~S8 of
the Supplementary Material gives the complete timing data and fits.

\section{Conclusions}
\label{sec:conclusions}
\label{sec:discussion}
In this work, we adapted the BUG scheme \cite{bug,bugadaptive} to update the pair basis
from the residual while solving the stationary MP2, dRPA and CCSD amplitude
equations. In the water-pentamer benchmarks, the adaptive basis closely
tracks the dominant subspace of the exact doubles amplitudes for both dRPA
and CCSD. For clusters of 5--30 waters, BUG-CCSD maintains a relative
correlation-energy error near 0.02\%, below the errors for fixed
perturbative pair bases of comparable rank \cite{hohenstein_gpu}. The projected Casida solve
selects the physical dRPA branch.
For alkane chains, water clusters, and Cu(111) slabs, the retained rank grows
approximately linearly with system size, giving cubic storage for the doubles factors.

Cancellation of component errors allows accurate interaction energies at
relatively loose truncation thresholds. We observe this for the dRPA
correlation contribution to CO adsorption on Cu(111) and for CCSD interaction
energies of benzene dimers. BUG-CCSD approaches the canonical DF-CCSD solver time
for the largest water clusters with a completed reference calculation and
becomes faster for the benzene dimer as the orbital basis grows.
Projecting the triples correction into the retained doubles basis reduces
its cost, but the larger interaction-energy errors in the benzene dimers
motivate a basis adapted to triples \cite{lesiuk_t,rickert_svd_t}.

Preserving the doubles factorization throughout the residual contractions
offers a route to lower formal scaling for BUG-CCSD. The CCSD update treats
the orbital-energy denominator through a Sylvester equation in the enlarged
pair basis, allowing the remaining contractions to use separable interaction
factors. The compressed
doubles also provide a starting point for the recently developed stochastic tensor contraction
\cite{stc}. A sampler that constructs importance weights directly from
the factors could avoid storing the full $O(N^4)$ importance-sampling
tensors, a major memory bottleneck of the stochastic approach.

\section*{Supplementary Material}
The Supplementary Material contains derivations, the projected (T)
formulation, additional benchmarks, timing data and regression checks.

\section*{Acknowledgments}
This work was supported by the Air Force Office of Scientific Research under
award FA9550-24-1-0096. C.V. acknowledges partial support from the Department
of Defense through the National Defense Science and Engineering Graduate
(NDSEG) Fellowship Program. We used GPT 6.1 Sol and 6-6.2 Astra to assist
with literature research, algorithm implementation, calculations, and manuscript
editing. We verified the code and results and take full responsibility for the
content of this work.

\section*{Data availability}
The plotted data, plotting scripts and calculation scripts for each figure are
available at
\url{https://github.com/nickirk/rank-adaptive-basis-update-and-galerkin-solvers-for-direct-rpa-and-ccsd-data}
(snapshot \texttt{3176ffc}). 

\section*{References}

% Apply the BibTeX style control without creating a numbered citation.
\makeatletter
\immediate\write\@auxout{\string\citation{aipBibliographyControl}}
\makeatother
\bibliographystyle{aipnum4-2}
\bibliography{references}

\end{document}